\documentstyle[multicol,aps,amsfonts,amssymb]{revtex}

\def \cB{{\cal B}}
\def \cC{{\cal C}}

\def \cG{{\cal G}}
\def \cH{{\cal H}}

\def \cK{{\cal K}}

\def \cS{{\cal S}}

\def \cX{{\cal X}}
\def \cY{{\cal Y}}

\def \defeq {:=}

\newcommand \set[1]{\{ #1 \}}
\newcommand \setcond [2]{\{ #1 \,|\, #2 \}}   

\newcommand \map [3] {#1 : #2 \rightarrow #3}
\newcommand \amap [3] {#1 : #2 \mapsto #3}
\newcommand \maps [2] {#1 \mapsto #2}

\newcommand \abs [1] {\left| #1  \right|}
\newcommand \norm [1] {\left \|  #1  \right \|}
\newcommand \trnorm [1] {\left \| #1  \right \|_1}
\newcommand \inorm [2] {\left \|  #1  \right \|_{#2}}
\newcommand \cbnorm [1] {\left \|  #1  \right \|_{\rm cb}}

\newcommand \tensprod [2]{{#1 \otimes #2}}

\newcommand \inv [1] {{#1}^{-1}}
\newcommand \id {\mathop{\rm id}\nolimits}

\newcommand\tr[1]{\mathop{{\rm tr}\,#1}\nolimits}
\def\idty{{\leavevmode{\rm 1\mkern -5.4mu I}}}

\newcommand{\braket}[2]{\langle #1 | #2 \rangle}
\newcommand{\ketbra}[2]{| #1 \rangle \langle #2 | }

\begin{document}

\draft

\title{Strictly contractive quantum channels \\
and physically realizable quantum computers}

\author{Maxim Raginsky\cite{author}}

\address{Center for Photonic Communication and Computing\\
Department of Electrical and Computer Engineering\\
Northwestern University, Evanston, IL 60208-3118}

\maketitle

\begin{abstract}

We study the robustness of quantum computers under the influence
of errors modelled by strictly contractive channels.  A channel
$T$ is defined to be strictly contractive if, for any pair of
density operators $\rho,\sigma$ in its domain, $\trnorm{T\rho -
T\sigma} \le k\trnorm{\rho-\sigma}$ for some $0 \le k < 1$ (here
$\trnorm{\cdot}$ denotes the trace norm).  In other words,
strictly contractive channels render the states of the computer
less distinguishable in the sense of quantum detection theory.
Starting from the premise that all experimental procedures can be
carried out with finite precision, we argue that there exists a
physically meaningful connection between strictly contractive
channels and errors in physically realizable quantum computers. We
show that, in the absence of error correction, sensitivity of
quantum memories and computers to strictly contractive errors
grows exponentially with storage time and computation time
respectively, and depends only on the constant $k$ and the
measurement precision.  We prove that strict contractivity rules
out the possibility of perfect error correction, and give an
argument that approximate error correction, which covers previous
work on fault-tolerant quantum computation as a special case, is
possible.

\end{abstract}

\begin{multicols}{2}

\section{Introduction}

Since it was first realized \cite{qcdiss} that maintaining
reliable operation of a large-scale (multiqubit) quantum computer
in the presence of environmental noise, as well as under the
combined influence of unavoidable imprecisions in state
preparation, manipulation, and measurement, will pose quite a
formidable obstacle to any experimental realization of the
computer \cite{note1}, many researchers have expended a
considerable effort devising various schemes for "stabilization of
quantum information." These schemes include, e.g., quantum
error-correcting codes (QECC's) \cite{kl}, noiseless quantum codes
(NQC's) \cite{zanras}, decoherence-free subspaces (DFS's)
\cite{dfs}, and noiseless subsystems (NS's) \cite{klvgennoise}.  (The last three of these schemes boil down to essentially the same thing, but are arrived at by different means.) However, each of these schemes relies for its efficacy upon explicit assumptions about the nature of the error mechanism.
Quantum error-correcting codes \cite{kl}, for instance, perform
best when different qubits in the computer are affected by
independent errors.  On the other hand, stabilization strategies
that are designed to handle collective errors
\cite{zanras,dfs,klvgennoise} make extensive use of various
symmetry arguments in order to demonstrate existence of the
so-called "noiseless subsystems," i.e., subsystems that are
effectively decoupled from the environment, even though the
computer as a whole certainly remains affected by errors.

In a recent publication \cite{zanardi}, Zanardi unified the
description of all above-mentioned schemes via a common algebraic
framework, thereby reducing the conditions for efficient
stabilization of quantum information to those based on symmetry
considerations.  The validity of this framework will ultimately be
decided by experiment, but it is also quite important to test its
applicability in a theoretical setting which would make as minimal
of an assumption as possible concerning the exact nature of the
error mechanism, and yet would serve as an abstract embodiment of
the concept of a physically realizable (i.e., nonideal) quantum
computer.

In this respect, the assumption of finite precision \cite{note2}
of all physically realizable state preparation, manipulation, and
registration procedures is particularly important, and can even be
treated as an empirical given.  This premise is general enough to
subsume (a) fundamental limitations imposed by the laws of quantum
physics (e.g., impossibility of reliable discrimination between
any two density operators with nonorthogonal supports), (b)
practical constraints imposed by the specific experimental setting
(e.g., impossibility of synthesizing any quantum state or any
quantum operation with arbitrary precision), and (c)
environment-induced noise.

As a rule, imprecisions in preparation and measurement procedures
will give rise to imprecisions in the building blocks of the
computer (gates) because the precision of any experimental
characterization of these gates will always be affected by the
precision of preparation and measurement steps involved in such
characterization.  Conversely, precision of quantum gates will
affect precision of measurements because the closeness of
conditional probability measures (say, in total variation norm
\cite{durrett}), conditioned on the gate used, is bounded above by
the closeness of any two quantum gates in question \cite{bernvaz}.

The central goal of this paper is to offer an argument that the
concept of a {\em strictly contractive} quantum channel yields a
natural (and very economical) embodiment of the above
finite-precision assumption.  Defining a suitable distance
function $d(\cdot,\cdot)$ on the set of density operators, we say
that a channel $T$ is strictly contractive (with respect to $d$)
if there exists some $k \in [0,1)$ such that, for any pair
$\rho,\sigma$ of density operators, we have the uniform estimate
\begin{equation}
d(T\rho,T\sigma) \le kd(\rho,\sigma).
\end{equation}
For instance, the much studied depolarizing channel is strictly
contractive (with respect to the trace-norm distance, to be
defined later). If we assume the dominant error mechanism of the
computer to be strictly contractive, then the constant $k$ can be
thought of as a quantitative measure of the computer's
(im)precision.  While this approach may certainly be criticized as
reductionist \cite{brody}, its merit lies in the fact that it
brings out many essential features of physically realizable
quantum computers without invoking more specific assumptions.

Let us give a "sneak preview" of what is coming up.  First of all,
we establish that the set of all strictly contractive channels on
a particular quantum system (computer) {\bf Q} is dense in the set
of all channels on {\bf Q}.  Since finite-precision measurements
cannot distinguish a dense subset from its closure \cite{meyer},
we draw the conclusion that strictly contractive quantum channels
(SCQC's) serve as a physically meaningful abstract model of errors
in physically realizable computers.  This conclusion is further
supported by the fact that, in the presence of a strictly
contractive error mechanism, the probability of correctly
discriminating between any two equiprobable quantum states is
bounded away from unity (or,  equivalently, no two density
operators in the image of a SCQC have orthogonal supports).  Next
we use a particularly important property of SCQC's, namely
existence and uniqueness of their fixed points, to obtain uniform
dimension-independent estimates of decoherence rates of noisy
quantum memories and computers.  We also take up the question of
possibility of error correction (stabilization).  In this regard,
we obtain a rather strong result that strictly contractive
channels admit no noiseless subsystems.  The proof of this claim
utilizes ideas from representation theory of operator algebras
\cite{bratrob,zhelob} and depends in an essential way on the
property of strict contractivity.

The paper is organized as follows.  In Section \ref{sec:prelims}
we introduce the necessary background on quantum states and
channels, as well as some relevant facts from operator theory.
Strictly contractive quantum channels are introduced in Section
\ref{sec:scqc}, where we show that the set of strictly contractive
channels is dense in the set of all channels.  Then, in Section
\ref{sec:qip}, we give an interpretation of strict contractivity
in the framework of optimum quantum hypothesis testing
(\ref{ssec:optqd}) and then use the fixed point theorem for
strictly contractive channels to obtain estimates on decoherence
rates of noisy quantum memories and computers
(\ref{ssec:noisyqmem}).  In \ref{ssec:stab}, we present the proof
of nonexistence of noiseless subsystems in the presence of SCQC's.
The possibility of approximate error correction is addressed in
\ref{ssec:approxec}. Finally, in Section \ref{sec:conclude}, we
present concluding remarks and outline some open questions and
directions for future research.

\section{Preliminaries}
\label{sec:prelims}

\subsection{On notation}
\label{ssec:notation}

In this paper we will adhere to the following notational
conventions. First of all, the operator adjoint to $X$ will be
denoted by $X^*$, as is usually done in mathematical physics
literature.  Secondly, density operators will be denoted by $\rho$
and $\sigma$ (with subscripts, whenever necessary).  We will use
capital Latin letters to denote all other operators;  whenever no
ambiguity may arise, the action of a mapping $X$ on a density
operator $\rho$ will be written as $X\rho$. Finally, the Pauli
matrices will be written as $\varsigma_i$, $i \in \set{1,2,3}$.

\subsection{States}
\label{ssec:states}

Let $\cH$ be a finite-dimensional Hilbert space associated with
the computer {\bf Q}, and let $\cB(\cH)$ be the algebra of all
bounded operators on $\cH$ [since $\dim \cH < \infty$, the
qualification "bounded" is patently unnecessary, but we will
retain the notation $\cB(\cH)$, following standard usage].  The
set
\begin{equation}
\cS(\cH) \defeq \setcond{\rho \in \cB(\cH)}{\rho \ge 0; \tr{\rho}
=1}
\end{equation}
is the set of all density operators (states) of {\bf Q}.  We can
define a few norms on $\cB(\cH)$; since $\cH$ is
finite-dimensional, all norm topologies on it are equivalent.
First, we have the operator norm
\begin{equation}
\norm{X} \defeq \sup_{\psi \in \cH;
\norm{\psi}=1}\norm{X\psi},\qquad \forall X \in \cB(\cH).
\end{equation}
We can also define the class of Schatten $p$-norms \cite{bhatia}.
For any $X \in \cB(\cH)$, we let $\abs{X} \defeq (X^* X)^{1/2}$,
so that
\begin{equation}
\inorm{X}{p} \defeq (\tr{\abs{X}^p})^{1/p},\qquad \forall X \in
\cB(\cH); p = 1,2,\ldots.
\end{equation}
The Schatten 1-norm is better known as the trace norm; in the case
$p=2$, we recover the Hilbert-Schmidt norm.  In fact, for any $X
\in \cB(\cH)$, $\norm{X}_p \rightarrow \norm{X}$ as $p \rightarrow
\infty$.  For this reason, we can identify the operator norm
$\norm{\cdot}$ with $\inorm{\cdot}{\infty}$. All these norms
possess a very important property of unitary invariance
\cite{bhatia}:  for any unitaries $U,V$ and any $X \in \cB(\cH)$,
we have
\begin{equation}
\inorm{UXV}{p} = \inorm{X}{p},\qquad p = 1,2,\ldots,\infty.
\label{eq:uinv}
\end{equation}

The trace norm can be given a natural interpretation as a distance
between density operators \cite{kitaev}.  First of all, for any
$\rho \in \cS(\cH)$, we have $\trnorm{\rho}=1$.  Of especial
importance is the fact that, for any pair $\rho,\sigma \in
\cS(\cH)$, the trace-norm distance $\trnorm{\rho-\sigma}$ achieves
its maximum value of 2 if and only if $\rho \sigma = 0$ (i.e., if
and only if $\rho$ and $\sigma$ have orthogonal supports).  In the
case of two pure states $\ketbra{\phi}{\phi}$ and
$\ketbra{\psi}{\psi}$, this condition reduces to
$\braket{\phi}{\psi}=0$, i.e., the corresponding state vectors
$\phi,\psi \in \cH$ are orthogonal.  As we will see in Section
\ref{sec:qip}, the trace-norm distance also figures prominently in
the framework of optimal quantum hypothesis testing.

To close our discussion of states, we give two important
characterizations of the trace-norm distance.  Let $X$ be a
selfadjoint operator.  Then we can write $X$ as a difference of
two positive operators with orthogonal supports:  $X = X_+ - X_-$,
where $X_\pm \defeq (\abs{X} \pm X)/2$.  This is referred to as
the orthogonal decomposition of $X$.  Then $\abs{X} = X_+ + X_-$,
and
\begin{equation}
\trnorm{X} = \trnorm{X_+} + \trnorm{X_-} \equiv \tr{X_+} +
\tr{X_-}.
\end{equation}
Now let $\rho,\sigma$ be a pair of density operators.  Writing
$\rho-\sigma = R_+ - R_-$, we observe that $\tr{(\rho-\sigma)}=0$
implies $\tr{R_+} = \tr{R_-}$, and hence
\begin{equation}
\trnorm{\rho-\sigma} = 2 \tr{R_+}.
\end{equation}
Another useful relation is
\begin{equation}
\trnorm{\rho-\sigma} = 2\max_{0 \le F \le \idty}
\tr{F(\rho-\sigma)}, \label{eq:trnormopdef}
\end{equation}
where the inequality $0 \le F \le \idty$ should be taken to mean
$F \ge 0$ and $\idty - F \ge 0$.  In fact, since the set of all
such $F$ is convex, the maximum of the linear functional in
Eq.~(\ref{eq:trnormopdef}) is attained on an extreme point of
$\setcond{F}{0 \le F \le \idty}$, namely on the projector $P_+$
defined by
\begin{equation}
P_+R_+ = R_+,\quad P_+R_- = 0.
\end{equation}

\subsection{Channels}
\label{ssec:channels}

In quantum theory, the reversible evolutions of a closed quantum
system correspond to the automorphisms $\cS(\cH) \rightarrow U\cS(\cH)U^*$ with a unitary $U$, i.e., for any $\rho \in \cS(\cH)$, we have
$\maps{\rho}{T_U\rho \defeq U\rho U^*}$. The map
$\map{T_U}{\cS(\cH)}{\cS(\cH)}$ is affine, trace-preserving, and
positive (we will call a map positive if it takes positive
operators to positive operators).  Since any affine map on density
operators can be uniquely extended to a linear map on selfadjoint
operators \cite{holevo}, we can take $T_U$ to be linear.  Most
importantly, $T_U$ is invertible with $\inv{T}_U\rho
\defeq U^*\rho U$.

The general irreversible evolution
$\map{T}{\cS(\cH)}{\cS(\cH)}$ of an open quantum system will no
longer be given by an automorphism $U \cdot U^*$. We must
accordingly modify the requirements imposed on $T$.  It is obvious
that we have to drop the invertibility condition, so that $T$ is
now a trace-preserving positive linear map on $\cS(\cH)$ (we can
then extend it, by linearity, to selfadjoint trace-class
operators).  However, positivity alone is not sufficient.  In
order for $T$ to represent a physically admissible evolution, it
must be {\em completely positive} \cite{kraus}, i.e., the map
$\tensprod{T}{\id_n}$, where $\id_n$ is the identity operator on
the space $M_n({\mathbb C})$ of $n \times n$ complex matrices, must be positive for all $n$.  Unitary evolutions
$T_U$ obviously satisfy all these requirements.  In fact, we can
also include such transformations as measurements into this
framework by requiring all admissible evolutions to be
trace-nonincreasing completely positive linear maps on $\cS(\cH)$,
i.e., for any $\rho \in \cS(\cH)$, we have $\tr{T\rho} \le
\tr{\rho}$, so that
\begin{equation}
\maps{\rho}{\frac{T\rho}{\tr{T\rho}}}.
\end{equation}
Then $\tr{T\rho}$ can be naturally interpreted as the conditional
probability of transformation $T$ occurring given that the system
is initially in the state $\rho$.  We will call any
trace-preserving completely positive linear map on $\cS(\cH)$ a
{\em channel}.

There are many useful structure theorems for completely positive
maps.  For instance, the Kraus representation theorem \cite{kraus}
states that, for any completely positive map
$\map{T}{\cB(\cH)}{\cB(\cH)}$, there exists a collection
$\set{K_i}$ of bounded operators such that
\begin{equation}
T(X) = \sum_i K_i X K^*_i,\qquad \forall X \in \cB(\cH).
\end{equation}
If $T$ is trace-nonincreasing, then we have the bound
\begin{equation}
\sum_i K^*_i K_i \le \idty,
\end{equation}
where equality is achieved if and only if $T$ is trace-preserving.
In addition, if $T$ is a {\em unital} channel, i.e., $T(\idty) =
\idty$, then we also have
\begin{equation}
\sum_i K_i K^*_i = \idty.
\end{equation}

Now we must adopt a suitable metric on the set of all completely
positive maps on $\cB(\cH)$.  One possible candidate is the metric induced by the operator norm,
\begin{equation}
\norm{T} \defeq \sup_{X \in \cB(\cH); \norm{X}=1} \norm{T(X)}.
\end{equation}
Unfortunately, the operator norm is rather ill-behaved
\cite{paulsen}:  it is not stable with respect to tensor products.
In particular, if $T$ is a completely positive map, then the norm
$\norm{\tensprod{T}{\id_n}}$ can in general increase with $n$.  A
good choice then is the metric induced by the {\em norm of
complete boundedness} \cite{paulsen} (or cb-norm), defined as
\begin{equation}
\cbnorm{T} \defeq \sup_n \norm{\tensprod{T}{\id_n}}.
\end{equation}
This norm has appeared, under different guises, in
Refs.~\onlinecite{kitaev}, \onlinecite{aharonov}, and
\onlinecite{gbcz}. For any selfadjoint trace-class operator $X$ on $\cH$ and any two maps $S,T$ on $\cB(\cH)$ with finite cb-norm (in the case of finite-dimensional $\cH$, this is always true \cite{paulsen}), we have the relations \cite{aharonov}
\begin{eqnarray}
\trnorm{T(X)} &\le & \cbnorm{T}\trnorm{X}, \label{eq:cbnorm1} \\
\cbnorm{TS} &\le & \cbnorm{T}\cbnorm{S}, \label{eq:cbnorm2} \\
\cbnorm{\tensprod{T}{S}} &=& \cbnorm{T}\cbnorm{S}.
\label{eq:cbnorm3}
\end{eqnarray}
Furthermore, for any channel $T$, we have \cite{holwer}
$\cbnorm{T} = 1$.

If two channels $T,S$ are close in cb-norm, then, for any density
operator $\rho$, the corresponding states $T\rho,S\rho$ are close
in trace norm since, from Eq.~(\ref{eq:cbnorm1}), it follows that
\begin{equation}
\trnorm{T\rho-S\rho} = \trnorm{(T-S)\rho} \le \cbnorm{T-S}.
\end{equation}
In fact, the above estimate cannot be loosened by adjoining a
second system with Hilbert space $\cK$ in some state $\sigma$,
entangling the two systems through some channel $K$ on
$\cS(\tensprod{\cH}{\cK})$, and then comparing the channels
$\tensprod{T}{R}$ and $\tensprod{S}{R}$, where $R$ is some
suitably chosen channel on $\cS(\cK)$. This is evident from the
estimate
\begin{equation}
\trnorm{ (\tensprod{T}{R}) K (\tensprod{\rho}{\sigma})-
(\tensprod{S}{R})K(\tensprod{\rho}{\sigma})} \le \cbnorm{T-S},
\label{eq:chdist}
\end{equation}
which can be easily obtained by repeated application of
Eqs.~(\ref{eq:cbnorm1})-(\ref{eq:cbnorm3}).  In other words, as far as the cb-norm distinguishability criterion is concerned, entangling the system with an auxiliary system will not improve distinguishability of the channels $T$ and $S$.  The cb-norm, however, is an extremely strong distinguishability measure: its definition already accounts for optimization with respect to entanglement and input states over Hilbert spaces of very large (but finite) dimension. There exist weaker measures of channel distinguishability (such as the channel fidelity \cite{raginsky}), which describe how channels may be distinguished with only finite resources.  Using these weaker criteria, one may show that entanglement does improve practical distinguishability of both states and channels \cite{dariano_entmeas}.

Before we go on, we must mention that, for the present purposes,
we only need to consider channels that map operators on some
Hilbert space $\cH$ to operators on the same Hilbert space.  In
general, this does not have to be true.  For instance, if the
Hilbert space in question is a tensor product
$\tensprod{\cH_1}{\cH_2}$, then the partial trace over $\cH_2$ can
be treated as a channel $\map{{\rm
tr}_2}{\cS(\tensprod{\cH_1}{\cH_2})}{\cS(\cH_1)}$.

\subsection{Some facts from operator theory}
\label{ssec:optheory}

We close Section \ref{sec:prelims} by listing some facts from
operator theory, which will be necessary in the sequel.  Let $\cX$
be a metric space with the corresponding metric $d(\cdot,\cdot)$.
An operator $\map{A}{\cX}{\cX}$ is called a {\em contraction} if,
for any $x,y \in \cX$, $d(Ax,Ay) \le d(x,y)$, and a {\em strict
contraction} if there exists some $k \in [0,1)$ such that
$d(Ax,Ay) \le kd(x,y)$.  If $\cX$ is a complete metric space, then
the contraction mapping principle \cite{reedsimon} states that any
strict contraction $A$ on $\cX$ has a unique fixed point.  In
other words, the problem $Ax = x$ has a unique solution on $\cX$.
If $\cY$ is a closed subset of $\cX$, then it follows that any
strict contraction $\map{A}{\cY}{\cY}$ has a unique fixed point on
$\cY$.

Strict contractivity is a remarkably strong property.  Indeed, if
we pick any $y \in \cY$, then the sequence of iterates $A^ny$
converges to the fixed point $y_0$ of $A$ exponentially fast,
because
\begin{equation}
d(A^ny,y_0) = d(A^ny,A^ny_0) \le k^nd(y,y_0). \label{eq:expconv}
\end{equation}
This fact is of tremendous use in numerical analysis when one
wants to solve a fixed-point problem $Ay=y$ via iteration method
with some initial guess $\hat{y}$.  If the operator $A$ is a
strict contraction on a closed subset of a complete metric space,
then, for any choice of $\hat{y}$, the iteration method is
guaranteed to zero in on the solution in $O(\log \epsilon^{-1})$
steps, where $\epsilon$ is the desired precision.

It should be noted that existence and uniqueness of a fixed point
of some operator $A$ are, by themselves, not sufficient to
guarantee convergence of the sequence of iterates $A^ny$ for any
point $y$ in the domain of $A$. Indeed, according to the
Leray-Schauder-Tychonoff theorem \cite{reedsimon}, any continuous
map on a compact convex subset of a locally convex space $\cX$ has
at least one fixed point.  Furthermore, any {\em weak contraction}
on a compact subset $\cC$ of a Banach space, i.e., a map
$\map{W}{\cC}{\cC}$ with the property $\norm{Wx-Wy} < \norm{x-y}$
for any $x,y \in \cC$, has a unique fixed point \cite{stakgold}.
The key to the rapid convergence in Eq.~(\ref{eq:expconv}) is the
fact that a strict contraction $\map{A}{\cY}{\cY}$ shrinks
distances between points of $\cY$ {\em uniformly}.

\section{Strictly contractive quantum channels}
\label{sec:scqc}

\subsection{Definiton and examples}
\label{ssec:defex}

Let $\cH$ be the finite-dimensional Hilbert space associated with
some quantum system {\bf Q}.  Then, as follows easily from
Eq.~(\ref{eq:cbnorm1}), any channel $T$ on $\cS(\cH)$ is a
contraction:
\begin{equation}
\trnorm{T\rho - T\sigma} \le \trnorm{\rho-\sigma},\qquad \forall
\rho,\sigma \in \cS(\cH).
\end{equation}
In other words, no channel can make any $\rho,\sigma$ more
distinguishable. For a channel $T$, we define the {\em
contractivity modulus}
\begin{equation}
\kappa(T) \defeq \sup_{\rho,\sigma \in \cS(\cH)} \frac{
\trnorm{T\rho-T\sigma}}{\trnorm{\rho-\sigma}}. \label{eq:contrmod}
\end{equation}
Any channel $T$ with $\kappa(T) < 1$ is {\em strictly
contractive}, and thus has a unique fixed point $\rho_T \in
\cS(\cH)$.

The depolarizing channel $D_p,0<p<1$, whose action on an arbitrary
$\rho \in \cS(\cH)$ is given by
\begin{equation}
D_p\rho \defeq p \frac{\idty}{d} + (1-p)\rho,
\label{eq:depolarize}
\end{equation}
where $d = \dim \cH$, is manifestly strictly contractive with
$\kappa(D_p) = 1-p$.  The maximally mixed state $\idty/d$ is the
unique fixed point of $D_p$ for any $p$.  When $d=2$, so that $\cH
= {\mathbb C}^2$, the action of $D_p$ on $\cS(\cH)$ can be
visualized as a uniform rescaling of the Bloch-Poincar\'e ball by
a factor of $\kappa(D_p)$, and the term "strictly contractive"
thus becomes especially apt.  It also turns out that, for any two
depolarizing channels $D_p$ and $D_q$, their tensor product is
also strictly contractive.  In order to show this, we use the fact
that a density operator on $\tensprod{{\mathbb C}^2}{{\mathbb
C}^2}$ can be written as \cite{mahweb}
\begin{eqnarray}
\rho &=& \frac{1}{4}\left(  \tensprod{\idty}{\idty} + \sum_k
\alpha_k
 \tensprod{\varsigma_k}{\idty} + \sum_k \beta_k
 \tensprod{\idty}{\varsigma_k} \right. \nonumber \\
 & + & \left. \sum_{k,l} \theta_{kl}
 \tensprod{\varsigma_k}{\varsigma_l}\right ),
\label{eq:bipartdo}
\end{eqnarray}
where the vectors
$\bbox{\alpha}\defeq(\alpha_1,\alpha_2,\alpha_3)$ and
$\bbox{\beta}\defeq(\beta_1,\beta_2,\beta_3)$ are the {\em
coherence vectors} of the first and second qubit respectively,
while the matrix $\Theta$ with entries $\theta_{kl}$ is called the
{\em correlation tensor} of $\rho$.  The action of the
depolarizing channel on an arbitrary operator $X \in \cB({\mathbb
C}^2)$ can be described as
\begin{equation}
D_p (X) = p (\tr{X}) \frac{\idty}{2} + (1-p)X
\end{equation}
[if $X$ is a density operator, this reduces to
Eq.~(\ref{eq:depolarize})].  Thus $D_p \varsigma_k =
(1-p)\varsigma_k$, which yields
\begin{eqnarray}
(\tensprod{D_p}{D_q})\rho &=&
\frac{1}{4}\left(\tensprod{\idty}{\idty} + (1-p)
\sum_k \alpha_k \tensprod{\varsigma_k}{\idty} \right. \nonumber \\
& + & (1-q)\sum_k \beta_k \tensprod{\idty}{\varsigma_k} \nonumber \\
& + & \left. (1-p)(1-q)
\sum_{k,l}\theta_{kl}\tensprod{\varsigma_k}{\varsigma_l} \right) .
\end{eqnarray}
It is then straightforward to verify that $T_{pq} \defeq
\tensprod{D_p}{D_q}$ is strictly contractive with $\kappa(T_{pq})
= \max[(1-p),(1-q)]$.  In particular, the channel
$\tensprod{D_p}{D_p}$ is strictly contractive with
$\kappa(\tensprod{D_p}{D_p}) = \kappa(D_p) = 1-p$.  Strict
contractivity of the product channel $\tensprod{D_p}{D_p}$
provides an alternate explanation of the fact that the use of
entanglement cannot improve distinguishability of classical signals
transmitted through the depolarizing channel \cite{bfmp}.

In fact, since the trace-norm distance between any two density
operators on ${\mathbb C}^2$ is just the Euclidean distance
between their Bloch-Poincar\'e vectors, any strictly contractive
channel on $\cS({\mathbb C}^2)$ can be pictured as a rescaling of
the Bloch-Poincar\'e ball (which may not be isotropic, as long as
the maximum of the scaling ratio over all directions is strictly
less than one), possibly followed by translation and rotation.  As
shown in Ref.~\onlinecite{rsw}, for any channel $T$ on
$\cB({\mathbb C}^2)$ [which is just the space $M_2({\mathbb C})$
of $2 \times 2$ complex matrices], there exist unitaries $U,V$ and
vectors ${\mathbf v},{\mathbf t} \in {\mathbb R}^3$ such that
\begin{equation}
T\rho = U\left[T_{{\mathbf v},{\mathbf t}}\left(V\rho
V^*\right)\right]U^*, \label{eq:rswrep}
\end{equation}
where the action of $T_{{\mathbf v},{\mathbf t}}$ is defined, with
respect to the basis
$\set{\idty,\varsigma_1,\varsigma_2,\varsigma_3}$, as
\begin{equation}
T_{{\mathbf v},{\mathbf t}}(w_0\idty + {\mathbf
w}\cdot\bbox{\varsigma}) \defeq w_0\idty + \left[{\mathbf t}+({\rm
diag\,}{\mathbf v}){\mathbf w}\right]\cdot\bbox{\varsigma}.
\end{equation}
Assuming that ${\mathbf v}$ and ${\mathbf t}$ are such that the
map $T$ is indeed a channel \cite{rsw}, we see that $T$ is
strictly contractive whenever $\max_{i \in \set{1,2,3}}\abs{v_i} <
1$.  In fact, the contractivity modulus of $T$ satisfies
\begin{equation}
\kappa(T) = \max_{i \in \set{1,2,3}} \abs{v_i}.
\end{equation}

If $T_1,T_2$ are unital strictly contractive channels on
$\cS({\mathbb C}^2)$, then the product channel
$\tensprod{T_1}{T_2}$ is strictly contractive on
$\cS(\tensprod{{\mathbb C}^2}{{\mathbb C}^2})$.  Given a
representation (\ref{eq:rswrep}) of a channel $T$ on $\cS({\mathbb
C}^2)$, we see that $T$ is unital if and only if ${\mathbf t}
\equiv 0$.  Specifically, for $T_1$ and $T_2$ we have
\begin{eqnarray}
T_1\rho &=& U_1[T_{{\mathbf v}_1,0}(V_1\rho V^*_1)]U^*_1, \\
T_2\rho &=& U_2[T_{{\mathbf v}_2,0}(V_2\rho V^*_2)]U^*_2.
\end{eqnarray}
Thus the action of the channel $\tensprod{T_1}{T_2}$ on a density
operator $\rho$ over $\tensprod{{\mathbb C}^2}{{\mathbb C}^2}$ is
a successive application of the unitary channel
$(\tensprod{V_1}{V_2})\cdot(\tensprod{V^*_1}{V^*_2})$, the
rescaling transformation $\tensprod{T_{{\mathbf
v}_1,0}}{T_{{\mathbf v}_2,0}}$, and the unitary channel
$(\tensprod{U_1}{U_2})\cdot (\tensprod{U^*_1}{U^*_2})$ to $\rho$.
By unitary invariance of the trace norm, we only need to consider
the effect of $\tensprod{T_{{\mathbf v}_1,0}}{T_{{\mathbf
v}_2,0}}$.  Writing $\rho$ in the form of Eq.~(\ref{eq:bipartdo}),
we obtain
\begin{eqnarray}
(\tensprod{T_{{\mathbf v}_1,0}}{T_{{\mathbf v}_2,0}}) \rho &=&
\frac{1}{4}\left(\tensprod{\idty}{\idty} + \sum_k v^{(1)}_k
\alpha_k
\tensprod{\varsigma_k}{\idty} \right. \nonumber \\
& + & \sum_k v^{(2)}_k \beta_k \tensprod{\idty}{\varsigma_k} \nonumber \\
& + & \left. \sum_{k,l}v^{(1)}_k v^{(2)}_l
\theta_{kl}\tensprod{\varsigma_k}{\varsigma_l} \right),
\end{eqnarray}
where $v^{(i)}_j, i \in \set{1,2}, j \in \set{1,2,3},$ denotes the
$j$th component of ${\mathbf v}_i$.  By inspection,
\begin{equation}
\kappa(\tensprod{T_1}{T_2}) = \max_{i,j}\abs{v^{(i)}_j}.
\end{equation}
If at least one of the channels $T_1$ and $T_2$ is not unital, the
tensor product channel $T = \tensprod{T_1}{T_2}$ may not be
strictly contractive, even if $T_1$ and $T_2$ are.  This stems
from the fact that, in this case, the effect of $T$ on the
correlation tensor $\Theta$ of $\rho$ is determined not only by
$T$, but also by $\rho$ through the coherence vectors
$\bbox{\alpha}$ and $\bbox{\beta}$.

It is quite easy to see that any unital strictly contractive
channel maps all states to mixed states.  Let $d = \dim \cH$.
Then, for any unit vector $\psi \in \cH$, we have
\begin{equation}
\trnorm{\ketbra{\psi}{\psi}-\frac{\idty}{d}} = \frac{2(d-1)}{d}
\end{equation}
(this can be readily proved by expanding $\idty$ with respect to
an orthonormal basis containing $\psi$).  Now suppose that $T$ is
a strictly contractive unital channel that maps
$\ketbra{\psi}{\psi}$ to some other pure state
$\ketbra{\phi}{\phi}$.  Then
\begin{equation}
\trnorm{T\ketbra{\psi}{\psi}-\frac{T\idty}{d}} =
\trnorm{\ketbra{\phi}{\phi}-\frac{\idty}{d}} = \frac{2(d-1)}{d}.
\end{equation}
Furthermore, we must also have
\begin{equation}
\trnorm{T\ketbra{\psi}{\psi}-\frac{T\idty}{d}} \le \kappa(T)
\trnorm{\ketbra{\psi}{\psi}-\frac{\idty}{d}}.
\end{equation}
Hence, $\kappa(T) \ge 1$, which is a contradiction, since
$\kappa(T) < 1$ for any strictly contractive channel.

We can show that there exist channels that are not strictly contractive,
and yet contain no pure states in their image. Let $T$ be an
arbitrary channel on $\cS(\cH)$. According to the
Leray-Schauder-Tychonoff theorem, $T$ has at least one fixed point
on $\cS(\cH)$. Let us adjoin another system with the associated
Hilbert space $\cK$.  Then the channel $\tensprod{T}{\id}$ on
$\cS(\tensprod{\cH}{\cK})$ cannot be strictly contractive because,
for any fixed point $\rho_T$ of $T$ and any $\sigma \in \cS(\cK)$,
the product density operator $\tensprod{\rho_T}{\sigma}$ is a
fixed point of $\tensprod{T}{\id}$.  The channel
$\tensprod{T}{\id}$ is not even weakly contractive, because it
preserves the trace-norm distance between any two of its fixed
points.  However, if the image of $\cS(\cH)$ under $T$ contains no
pure states, then the image of $\cS(\tensprod{\cH}{\cK})$ under
$\tensprod{T}{\id}$ contains no pure states either because of the
relation \cite{ahw}
\begin{equation}
\inf_{\rho \in \cS(\cH)}S(T\rho) = \inf_{\rho \in
\cS(\tensprod{\cH}{\cK})}S\left((\tensprod{T}{\id})\rho\right),
\end{equation}
where $S(\rho)$ is the von Neumann entropy of the state $\rho$.

\subsection{Strictly contractive channels are dense in the set of all channels}

As we have mentioned in the Introduction, finite-precision
measurements cannot distinguish a dense subset from its closure.
Let us make this statement more precise.  Suppose we are presented
with some quantum system {\bf Q} in an unknown state $\rho$, and
we are trying to estimate the state.  Any physically realizable
apparatus will have finite resolution $\epsilon$, so that all
states $\rho'$ with $\trnorm{\rho-\rho'} < \epsilon$ are
considered indistinguishable from $\rho$.  Now, if $\cH$ is the
Hilbert space associated with {\bf Q}, and if $\Sigma$ is a dense
subset of $\cS(\cH)$, then, by definition of a dense subset, for
any $\epsilon > 0$ and any $\rho \in \cS(\cH)$, there will always
be some $\sigma \in \Sigma$ such that $\trnorm{\rho-\sigma} <
\epsilon$.

The same reasoning also applies to distinguishability of quantum
channels, except now the appropriate measure of closeness is
furnished by the cb-norm. Thus, if an experiment utilizes some
apparatus with resolution $\epsilon$, then any two channels $T,S$
with $\cbnorm{T-S} < \epsilon$ are considered indistinguishable
from each other.  There is, however, no fundamental difference
between distinguishability of states and channels because any
experiment purporting to distinguish any two channels $T$ and $S$
consists in preparing the apparatus in some state $\rho$ and then
making some measurements that would tell the states $T\rho$ and
$S\rho$ apart from each other.  Then, since for any state $\rho$,
$\trnorm{T\rho-S\rho} \le \cbnorm{T-S}$, the resolving power of
the apparatus that will distinguish between $T$ and $S$ is limited
by the resolving power of the apparatus that will distinguish
between $T\rho$ and $S\rho$.

In this regard, we have the following

\noindent{{\bf Proposition 1}.  Let $C(\cH)$ be the set of all
channels on $\cS(\cH)$, where $\cH$ is the Hilbert space
associated with the system {\bf Q}.  Then the set $C_{\rm
sc}(\cH)$ of all strictly contractive channels on $\cS(\cH)$ is a
$\cbnorm{\cdot}$-dense convex subset of $C(\cH)$.}

\noindent{{\bf Proof}.  We show convexity first.  Suppose $T_1,T_2
\in C_{\rm sc}(\cH)$.  Form the channel $S \defeq \lambda T_1 +
(1-\lambda)T_2$, $0 < \lambda < 1$.  Then, for any $\rho,\sigma
\in \cS(\cH)$, we have the estimate
\begin{eqnarray}
\trnorm{S\rho-S\sigma} & \le &  \lambda \trnorm{T_1\rho-T_1\sigma}
+
(1-\lambda) \trnorm{T_2\rho-T_2\sigma} \nonumber \\
& \le & \left[ \lambda \kappa(T_1) + (1-\lambda)\kappa(T_2)\right]
\trnorm{\rho-\sigma}.
\end{eqnarray}
Defining $\kappa \defeq \max\left[\kappa(T_1),\kappa(T_2)\right]$,
we get
\begin{equation}
\trnorm{S\rho-S\sigma} \le \kappa \trnorm{\rho-\sigma}.
\end{equation}
Since $T_1,T_2$ are strictly contractive, $\kappa < 1$, and
therefore $S \in C_{\rm sc}(\cH)$.} To prove density, let us fix
some $\sigma \in \cS(\cH)$.  Now the map $\amap{K_\sigma}{\rho \in
\cS(\cH)}{\sigma}$ is obviously a channel, which is furthermore
trivially strictly contractive because it maps all density
operators $\rho$ to $\sigma$.  Given $\epsilon > 0$, pick some
positive $n$ such that $1/n < \epsilon$.  For any $T \in C(\cH)$,
define
\begin{equation}
T_n \defeq \frac{1}{2n}K_\sigma + \left(1-\frac{1}{2n}\right)T.
\end{equation}
Clearly, $T_n \in C_{\rm sc}(\cH)$, and the estimate
\begin{equation}
\cbnorm{T-T_n} = \frac{1}{2n}\cbnorm{T-K_\sigma} \le \frac{1}{n} <
\epsilon
\end{equation}
finishes the proof.\hfill $\blacksquare$

This proposition indicates that, as far as physically realizable
(finite-precision) measurements go, there is no way to distinguish
any channel $T$ from some strictly contractive $T'$ with
$\cbnorm{T-T'} < \epsilon$, where $\epsilon$ is the resolution of
the measuring apparatus.  In this regard, it is interesting to
mention that any channel $T$ with $\cbnorm{T-\id} < \epsilon$ (for
some sufficiently small $\epsilon > 0$) cannot be distinguished
from a depolarizing channel.  Indeed, let $M$ be the channel that
maps all density operators $\rho$ to the maximally mixed state
$\idty/d$, where $d = \dim \cH$. Then it suffices to pick some
\begin{equation}
n > \frac{ \cbnorm{M-\id} } {\epsilon - \cbnorm{T-\id} },
\end{equation}
so that
\begin{equation}
\cbnorm{T - D_{1/n}} \le \cbnorm{T-\id} + (1/n)\cbnorm{M-\id} <
\epsilon.
\end{equation}

We note that a convex combination of any channel with a strictly
contractive channel is a strictly contractive channel.  Let $T \in
C$ be an arbitrary channel, and suppose that $T' \in C_{\rm sc}$
[from now on, we will not mention the Hilbert space $\cH$ when
talking about channels on $\cS(\cH)$, unless this omission might
cause ambiguity]. Define, for some $0 < \lambda <1$, the channel
$S \defeq \lambda T + (1-\lambda)T'$.  Then
\begin{eqnarray}
\trnorm{S\rho - S\sigma} & \le & \lambda \trnorm{T\rho-T\sigma} +
(1-\lambda)
\trnorm{T'\rho - T'\sigma} \nonumber \\
& \le & \left[\lambda + (1-\lambda)\kappa(T')\right]
\trnorm{\rho-\sigma}.
\end{eqnarray}
Since $\lambda + (1-\lambda)\kappa(T') < 1$, we conclude that $S
\in C_{\rm sc}$.

Finally, we mention that Proposition 1 implies that the set
$C_{\rm sc}^\idty$ of all unital strictly contractive channels is
a dense convex subset of the set $C^\idty$ of all unital channels.

\section{Implications for quantum information processing}
\label{sec:qip}

\subsection{Optimum quantum decision strategies}
\label{ssec:optqd}

In this subsection we explore an interesting connection between
the contractivity modulus of a channel and quantum detection
theory \cite{helstrom}. The archetypal problem in quantum
detection theory is that of optimum $M$-ary detection.  A quantum
system is prepared in a state $\rho$, drawn from a collection
$\set{\rho_i}^M_{i=1}$ of $M$ density operators, where $\rho_i$ is
selected with probability $\pi_i$. Our task is to determine, as
accurately as possible, which state $\rho_i$ has been drawn.  On
this system we can perform a measurement described by a positive
operator-valued measure (POVM), i.e., a collection
$\set{F_i}^M_{i=1}$ of operators that satisfy
\begin{eqnarray}
0 \le F_i &\le & \idty,\qquad i = 1,\ldots,M, \label{eq:povmcond1} \\
\sum^M_{i=1}F_i &=& \idty. \label{eq:povmcond2}
\end{eqnarray}
We seek a POVM that would solve the optimization problem
\begin{equation}
\bar{P}_c = \max_{\set{F_i}} \sum^M_{i=1}\pi_i \tr{F_i\rho_i},
\label{eq:maryqdet}
\end{equation}
subject to the constraints (\ref{eq:povmcond1}) and
(\ref{eq:povmcond2}).  The quantity being maximized in
(\ref{eq:maryqdet}) is the probability of correct decision using
the POVM $\set{F_i}$.  We will only consider the case $M=2$,
wherein the system can be in the state $\rho_1$ with probability
$\pi_1$, or in the state $\rho_2$ with probability $\pi_2 \equiv
1-\pi_1$.  In this case, we are considering two-element POVM's
$\set{F,\idty-F}$ with $0 \le F \le \idty$, and the optimization
problem (\ref{eq:maryqdet}) takes the form
\begin{equation}
\bar{P}_c = \max_{0 \le F \le \idty} \left[ \pi_1 \tr{F\rho_1} +
\pi_2 \tr{(\idty-F)\rho_2} \right],
\end{equation}
or, equivalently,
\begin{equation}
\bar{P}_c = \pi_2 + \max_{0 \le F \le \idty} \tr{\left[ F (\pi_1
\rho_1 - \pi_2 \rho_2) \right]}. \label{eq:optbinqdet}
\end{equation}
We interpret $\tr{F\rho_1}$ as the conditional probability that
the measurement using the POVM $\set{F,\idty-F}$ correctly
determines the state of the system to be $\rho_1$; similarly,
$\tr{(\idty-F)\rho_2}$ is the conditional probability that the
state $\rho_2$ is identified correctly.  Then $\tr{F\rho_2}$ and
$\tr{(\idty-F)\rho_1}$ respectively are the conditional
probabilities of mistaking $\rho_2$ for $\rho_1$ and vice versa.

We can easily show that
\begin{equation}
\bar{P}_c = \frac{1}{2} + \frac{1}{2}\trnorm{\pi_1\rho_1 -
\pi_2\rho_2}. \label{eq:optbinprob}
\end{equation}
Writing down the orthogonal decomposition $\pi_1\rho_1 -
\pi_2\rho_2 = R_+ - R_-$, we get $\tr{R_+} = \pi_1 - \pi_2 +
\tr{R_-}$.  Now
\begin{equation}
\max_{0 \le F \le \idty} \tr{F(R_+ - R_-)} = \tr{R_+},
\end{equation}
where the maximum is attained by choosing $F$ to be a projection
operator with $FR_+ = R_+$ and $FR_- = 0$. Since
\begin{equation}
\tr{\abs{\pi_1\rho_1-\pi_2\rho_2}} = \tr{R_+} + \tr{R_-} =
2\tr{R_+} + \pi_2 - \pi_1,
\end{equation}
we finally arrive at Eq.~(\ref{eq:optbinprob}), which clearly
exhibits the role of the trace-norm distance in optimum quantum
hypothesis testing.  It can be proved \cite{yuen} that $\bar{P}_c
= 1$ if and only if $\rho_1 \rho_2 = 0$, in which case
$\trnorm{\pi_1\rho_1 - \pi_2\rho_2} = \pi_1+\pi_2 = 1$.

Now suppose that the state of the system is given by one of two
equiprobable density operators $\rho_1,\rho_2$. Suppose
furthermore that $\rho_1\rho_2 = 0$, so that
$\trnorm{\rho_1-\rho_2} = 2$.  Then there exists a measurement
that would correctly distinguish between $\rho_1$ and $\rho_2$
with probability one. Since any channel $T$ will generally
decrease the trace-norm distance $\trnorm{\rho_1-\rho_2}$, it can
happen that the states $T\rho_1$ and $T\rho_2$ no longer have
orthogonal supports, and thus the optimum decision strategy will
fail with nonzero probability $P_e \equiv 1 - \bar{P}_c$.

If $T$ is a weakly contractive channel, then no two density
operators in its image have orthogonal supports, but the
probability of error $P_e$ can, in principle, be made arbitrarily
small.  If, however, $T$ is strictly contractive, then we have the
trivial, but important,

\noindent{{\bf Lemma 1}.  Let $T$ be a strictly contractive
channel.  Then, for any pair $\rho_1,\rho_2$ of equiprobable
density operators, the optimum decision strategy for $T\rho_1$ and
$T\rho_2$ is such that
\begin{equation}
\bar{P}_c \le \frac{1+\kappa(T)}{2} < 1. \label{eq:lemma1}
\end{equation}
} \noindent{{\bf Proof}.  Fix a pair $\rho_1,\rho_2$ of density
operators.  Then, using Eq.~(\ref{eq:optbinprob}), we get
\begin{equation}
\bar{P}_c = \frac{1}{2}+\frac{1}{4}\trnorm{T\rho_1-T\rho_2} \le
\frac{1}{2} + \frac{\kappa(T)}{4}\trnorm{\rho_1-\rho_2}.
\end{equation}
Since $\trnorm{\rho_1-\rho_2} \le 2$, we obtain
Eq.~(\ref{eq:lemma1}).\hfill $\blacksquare$}

We note that the statement of the above lemma can be extended to
general channels.  For instance, if $T$ is a channel with the
property that there exists at least one pair $\rho,\sigma$ of
density operators such that $\trnorm{T\rho-T\sigma} =
\trnorm{\rho-\sigma}$, then the bound (\ref{eq:lemma1}) is
obviously
\begin{equation}
\bar{P}_c \le 1. \label{eq:modlemma1}
\end{equation}
If $T$ is a weakly contractive channel, then the inequality
(\ref{eq:modlemma1}) becomes strict, but $\bar{P}_c$ can, at least
in principle, be made arbitrarily close to one. This is decidedly
not the case for a strictly contractive channel, in which case
Lemma 1 states that for any pair of equiprobable density operators
$\rho,\sigma$, the probability $\bar{P}_c$ of correctly
discriminating between them is bounded away from one.

The discussion in this subsection lends further support to our
argument that strictly contractive channels serve as an abstract
model of errors in physically realizable quantum computers.  In
any realistic setting, no event occurs with probability exactly
equal to unity.  For instance, we can never prepare a pure state
$\ketbra{\psi}{\psi}$, but rather a mixture
$(1-\epsilon)\ketbra{\psi}{\psi} + \epsilon \rho$, where both
$\epsilon$ and $\rho$ depend on the particulars of the preparation
procedure.  Similarly, the measuring device that would ideally
identify $\ketbra{\psi}{\psi}$ perfectly will instead be realized
by $(1-\delta)\ketbra{\psi}{\psi} + \delta F$, where $\delta$ and
the operator $F, 0 \le F \le \idty$, are again determined by
practice.  If we assume that, in any physically realizable
computer, all state preparation, manipulation, and registration
procedures can be carried out with finite precision, then it is
reasonable to expect that there exist strict bounds on all
probabilities that figure in the description of the computer's
operation.

\subsection{Decoherence rates of noisy quantum memories and computers}
\label{ssec:noisyqmem}

So far, we have established two important properties of strictly
contractive channels.  Firstly, any channel $T$ can be
approximated, in cb-norm, by a strictly contractive channel $T'$,
and there will always be some finite-precision measurement which
will not be able to distinguish $T$ from $T'$.  Secondly, any
measurement that would, in principle, distinguish some pair
$\rho,\sigma$ of density operators with certainty, will fail with
probability at least $[1-\kappa(T)]/2$ in the presence of a
strictly contractive error channel $T$.  The latter statement can
also be phrased as follows:  no two density operators in the image
$T\cS(\cH)$ of $\cS(\cH)$ under some $T \in C_{\rm sc}$ have
orthogonal supports; furthermore, the trace-norm distance between
any two density operators in $T\cS(\cH)$ is bounded from above by
$2\kappa(T)$.

In this subsection, we obtain dimension-independent estimates on
decoherence rates of quantum memories and computers under the
influence of strictly contractive noise and without any error
correction (the possibility of error correction will be addressed
in \ref{ssec:stab} and \ref{ssec:approxec}).

We treat quantum memories (registers) first.  Suppose that we want
to store some state $\rho_0 \in \cS(\cH)$ for time $t$ in the
presence of errors modelled by some strictly contractive channel
$T$.  Let $\tau$ be the decoherence timescale, with $\tau \ll t$,
and let $n = \lceil t/\tau \rceil$. The final state of the
register is then $\rho_n = T^n\rho_0$. If $\rho_T$ is the unique
fixed point of $T$, then
\begin{equation}
\trnorm{\rho_n - \rho_T} = \trnorm{T^n\rho_0 - T^n\rho_T} \le
\kappa(T)^n \trnorm{\rho_0 - \rho_T}. \label{eq:qmem}
\end{equation}
In other words, the state $\rho_0$, stored in a quantum register
in the presence of strictly contractive noise $T$, evolves to the
fixed state $\rho_T$ of $T$, and the convergence is incredibly
rapid.  Let us consider a numerical example.  Suppose that
$\kappa(T) = 0.9$, and that initially the states $\rho_0$ and
$\rho_T$ have orthogonal supports, so $\trnorm{\rho_0 - \rho_T} =
2$.  Then, after $n=10$ iterations (i.e., $t=10\tau$), we have
$\trnorm{\rho_n - \rho_T} \le 0.697$, and the probability of
correct discrimination between $\rho_n$ and $\rho_T$ is only
0.674.  Note that the decoherence rate estimate
\begin{equation}
r(n;\rho,T) \defeq \frac{\trnorm{\rho_n -
\rho_T}}{\trnorm{\rho_0-\rho_T}} \le \kappa(T)^n \label{eq:decmem}
\end{equation}
does not depend on the dimension of $\cH$, but only on the
contractivity modulus $\kappa(T)$ and on the relative storage
duration $n$.  In other words, quantum registers of {\em any} size
are equally sensitive to strictly contractive errors.

Obtaining estimates on decoherence rates of computers is not so
simple because, in general, the sequence $\set{\rho_n}$, where
$\rho_n$ is the overall state of the computer after $n$
computational steps, does not have to be convergent.  Let us first
fix the model of a quantum computer.  We define \cite{note3} an
{\em ideal quantum circuit of size $n$} to be an ordered $n$-tuple
of unitaries $U_i$, where each $U_i$ is a tensor product of
elements of some set $\cG$ of universal gates \cite{qgates}. The
set $\cG$ will in, general, be a dense subgroup of the group
$U(\cH)$ of all unitary operators on $\cH$.  For some error
channel $T$, a {\em $T$-noisy quantum circuit of size $n$ with $k$
error locations} is an ordered $(n+k)$-tuple containing $n$
channels $\hat{U}_i \defeq U_i \cdot U^*_i$, where the unitaries
$U_i$ are of the form described above, as well as $k$ instances of
$T$.  We will assume, for simplicity, that each $T$ is preceded
and followed by some $\hat{U}_i$ and $\hat{U}_{i+1}$ respectively.
Based on this definition, the ``noisiest'' computer for fixed $T$
and $n$ is modelled by a $T$-noisy quantum circuit of size $n$
with $n$ error locations, i.e., a $2n$-tuple of the form
$(\hat{U}_1,T,\hat{U}_2,T,\ldots,\hat{U}_n,T)$.  If the initial
state of the computer is $\rho_0$, then we will use the notation
\begin{equation}
\rho_n = \left(\prod^n_{i=1}T\hat{U}_i\right)\rho_0
\label{eq:ncomp}
\end{equation}
to signify the state of the computer after $n$ computational
steps. In the above expression, the product sign should be
understood in the sense of composition $T \circ \hat{U}_n \circ
\ldots \circ T \circ \hat{U}_1$.

Given an arbitrary sequence of computational steps, the sequence
$\set{\rho_n}$, defined by Eq.~(\ref{eq:ncomp}) (assuming that $n$
is suficiently large, i.e., the computation is sufficiently long)
need not be convergent.  However, if the channel $T$ is strictly
contractive, then, for any $\epsilon > 0$, there exists some $N_0$
such that, for any pair of initial states $\rho_0,\sigma_0 \in
\cS(\cH)$, the states $\rho_n,\sigma_n$, $n \ge N_0$, will be
indistinguishable from each other.  In other words, any two
sufficiently lengthy computations will yield nearly the same final state.
Using Eq.~(\ref{eq:ncomp}), as well as unitary invariance of the
trace norm, we obtain
\begin{eqnarray}
\trnorm{\rho_n - \sigma_n} &=&
\trnorm{\left(\prod^n_{i=1}T\hat{U}_i\right) (\rho_0 - \sigma_0)}
\nonumber \\
& \le & \kappa(T)^n \trnorm{\rho_0 - \sigma_0}. \label{eq:deccomp}
\end{eqnarray}
Now suppose that, at the end of the computation, we perform a
measurement with precision $\epsilon$, i.e., any two states
$\rho,\sigma$ with $\trnorm{\rho-\sigma} < \epsilon$ are
considered indistinguishable.  Then, if the computation takes at
least $N_0 = \lceil \log(\epsilon/2)/\log \kappa(T) \rceil$ steps,
we will have $\trnorm{\rho_n - \sigma_n} < \epsilon$ for all $n
\ge N_0$.  For a numerical illustration, we take $\kappa(T) = 0.9$
and $\epsilon = 0.01$, which yields $N_0 = 50$.  In other words,
the result of any computation that takes more than 50 steps in the
presence of a strictly contractive channel $T$ with
$\kappa(T)=0.9$ is untrustworthy since we will not be able to
distinguish between any two states $\rho$ and $\sigma$ with
$\trnorm{\rho-\sigma} < 0.01$. Again, $N_0$ depends only on the
contractivity modulus of $T$ and on the measurement precision
$\epsilon$, not on the dimension of $\cH$, at least not
explicitly.  We note that, if the state of the computer is a
density operator over a $2^k$-dimensional Hilbert space, then any
efficient quantum computation will take $O({\rm Poly}(k))$ steps,
and therefore the sensitivity of the computer to errors grows
exponentially with $k$.

Let us consider some cases where the sequence $\set{\rho_n}$ does
converge. Suppose first that the channel $T \in C_{\rm sc}$ is
unital.  Then, since each channel $\hat{U}_i$ is unital as well,
the sequence $\set{\rho_n}$ converges exponentially fast to the
maximally mixed state $\idty/d$, where $d = \dim \cH$. If the computation employs a static algorithm, i.e.,
$\hat{U}_i = \hat{U}$ for all $i$ (this is true, e.g., in the case
of Grover's search algorithm \cite{grover}), then the channel $S
\defeq T\hat{U}$ is also strictly contractive, and $\kappa(S) =
\kappa(T)$ by unitary invariance of the trace norm.  Denoting the
fixed point of $S$ by $\rho_S$, we then have
\begin{equation}
\trnorm{\rho_n - \rho_S} = \trnorm{S^n\rho_0 - S^n\rho_S} \le
\kappa(T)^n \trnorm{\rho_n - \rho_S},
\end{equation}
i.e., the output state of any sufficiently lengthy computation
with a static algorithm will be indistinguishable from the fixed
point $\rho_S$ of $S = T\hat{U}$.

\subsection{Impossibility of perfect error correction}
\label{ssec:stab}

After we have seen in the previous subsection that quantum
memories and computers are ultrasensitive to errors modelled by
strictly contractive channels, we must address the issue of error
correction (stabilization of quantum information). Since we have
not made any specific assumptions (beyond strict contractivity)
about the errors affecting the computer, it is especially
important to investigate the possibility of error correction, if
only to determine the limitations on the robustness of physically
realizable quantum computers from the foundational standpoint.

First of all, strict contractivity rules out the possibility of
perfect quantum error-correcting codes \cite{kl}.  Let us recall
the basics of QECC's.  We seek to protect a quantum system with a
$k$-dimensional Hilbert $\cH$ space by realizing it as a subspace
$\cK$ (called the {\em code}) of a larger $n$-dimensinal Hilbert
space $\cH_c$, known as the {\em coding space}.  In other words,
the Hilbert space $\cH$ is embedded in the coding space $\cH_c$
via the isometric encoding operator $\map{V_{\rm enc}}{\cH}{\cK}$.
Now, for any channel $T$ on $\cS(\cH_c)$, a theorem of Knill and
Laflamme \cite{kl} asserts that a subspace $\cK$ of $\cH_c$ can
serve as a {\em $T$-corrrecting code} if and only if there exists
some channel $S$ on $\cS(\cH_c)$ such that $\left. ST
\right|_{\cK} = \id$.  In other words, $S$ is the left inverse of
the restriction of $T$ to $\cS(\cK)$. However, if the channel $T$ on
$\cS(\cH_c)$ is strictly contractive, then no subspace $\cK$ of
$\cH_c$ is a $T$-correcting code. Suppose, to the contrary, that
such a subspace $\cK$ exists, and let $\set{e_\mu}$ be any
orthonormal basis of $\cK$. Then there also exists some channel
$S$ on $\cS(\cH_c)$ that satisfies the Knill-Laflamme condition
for $T$ and $\cK$.  Thus
\begin{equation}
\trnorm{ST(\ketbra{e_\mu}{e_\mu}-\ketbra{e_\nu}{e_\nu})} =
\trnorm{\ketbra{e_\mu}{e_\mu}-\ketbra{e_\nu}{e_\nu}}
\end{equation}
for all $\mu,\nu$.  But, using Eq.~(\ref{eq:cbnorm1}) and strict
contractivity of $T$, we also have
\begin{equation}
\trnorm{\ketbra{e_\mu}{e_\mu} - \ketbra{e_\nu}{e_\nu}} \le
\kappa(T)\trnorm{\ketbra{e_\mu}{e_\mu}-\ketbra{e_\nu}{e_\nu}},
\end{equation}
which yields $\kappa(T) \equiv 1$.  Since $T$ is strictly
contractive, this is a contradiction, and therefore no subspace
$\cK$ of $\cH_c$ is a $T$-correcting code.

Before we go on, we must mention that the Knill-Laflamme theorem
provides also for approximately correctable channels.  That is,
let $\set{K_i}$ be the set of the Kraus operators of some channel
$T$ on $\cS(\cH_c)$.  For any subset $\Lambda$ of $\set{K_i}$, we
can define the completely positive map $T_\Lambda$ via
\begin{equation}
T_\Lambda(X) \defeq \sum_{K_i \in \Lambda}K_i X K^*_i,\quad
\forall X \in \cB(\cH_c).
\end{equation}
Then a subpace $\cK$ of $\cH_c$ can serve as a
$T_\Lambda$-correcting code if there exists some channel $S$ on
$\cS(\cH_c)$ such that $\left. ST_\Lambda \right|_{\cK} \propto
\id$.  If $\cbnorm{T-T_\Lambda}$ is sufficiently small, then the
errors modelled by the channel $T$ are approximately correctable.
Thus, in and of itself, the impossibility of perfect error
correction for strictly contractive channels is not likely to be a
serious problem.

However, strict contractivity also proscribes the existence of
noiseless subsystems in the sense of Knill-Laflamme-Viola \cite{klvgennoise} and Zanardi \cite{zanardi}, the
essence of which we now summarize.  Given some quantum system
(computer) {\bf Q} with the associated finite-dimensional Hilbert
space $\cH$, we consider the error channel $T$ with Kraus
operators $K_i$.  We define the {\em interaction algebra
$\frak{K}$ of $T$} as a *-algebra generated by $K_i$.  It is
obvious that $\cK$ is an algebra with identity because of the
condition $\sum_i K^*_i K_i = \idty$.  However, since the Kraus
representation of a channel $T$ is not unique, we must make sure
that, for any two choices $\set{K_i}$ and $\set{K_\mu}$ of Kraus
representations of $T$, the corresponding interaction algebras are
equal.  Using the fact that any two Kraus representations of a
channel are connected via
\begin{equation}
K_i = \sum_\mu v_{i\mu}K_\mu,
\end{equation}
where $v_{i\mu}$ are the entries of a matrix $V$ with $V^*V =
\idty$ (assuming that one of the sets $\set{K_i}$ and
$\set{K_\mu}$ is padded with zero operators in order to ensure
that they have the same cardinality), we see that it is indeed the
case that the interaction algebra of a channel $T$ does not depend
on the particular choice of the Kraus operators.

The existence of noiseless subsystems of {\bf Q} with respect to
$T$ hinges on the reducibility of the interaction algebra
$\frak{K}$. Since $\frak{K}$ is a uniformly closed
\protect{*-subalgebra} of $\cB(\cH)$, it is a finite-dimensional
C*-algebra, and is therefore isomorphic to a direct sum of $r$
full matrix algebras, each of which appears with multiplicity
$m_i$ and has dimension $n^2_i$ (i.e., it is an algebra of $n_i
\times n_i$ complex matrices).  Thus $\dim {\frak K} =
\sum^r_{i=1}n^2_i$. The {\em commutant} $\frak{K}'$ of $\frak{K}$
is defined as the set of all operators $X \in \cB(\cH)$ that
commute with all $K \in \frak{K}$. From the Wedderburn theorem
\cite{zhelob} it follows that each $K \in \frak{K}$ has the form
\begin{equation}
K = \bigoplus^r_{i=1} \tensprod{\idty_{m_i}}{K_i},\qquad K_i \in
M_{n_i}(\mathbb{C}), \label{eq:intalg1}
\end{equation}
and that each $K' \in {\frak K}'$ has the form
\begin{equation}
K' = \bigoplus^r_{i=1} \tensprod{K^\prime_i}{\idty_{n_i}},\qquad
K^\prime_i \in M_{m_i}(\mathbb{C}). \label{eq:intalg2}
\end{equation}
Thus $\dim {\frak K}' = \sum^r_{i=1}m^2_i$.  We have the
corresponding isomorphism
\begin{equation}
\cH \simeq \bigoplus^r_{i=1} \tensprod{{\mathbb C}^{m_i}}{{\mathbb
C}^{n_i}}, \label{eq:hsisomorph}
\end{equation}
and each factor ${\mathbb C}^{m_i}$ is referred to as a {\em
noiseless subsystem} because it is effectively decoupled from the
error channel $T$.  It is rather obvious that, in order to be of
any use, a noiseless subsystem must be nontrivial, i.e., at least
two-dimensional.  Now, if the interaction algebra ${\frak K}$ is
irreducible, then $\dim {\frak K}' = 1$, and no noiseless
subsystems exist.  There is a simple criterion of irreducibility
of an algebra, the Schur's lemma \cite{bratrob}, which states that
a *-algebra $\frak{A}$ is irreducible if and only if its commutant
$\frak{A}'$ consists of complex multiples of the identity.  We are
now ready to state two main results of this subsection.

\noindent{{\bf Proposition 2}.  Let $T$ be a strictly contractive
unital channel.  Then $T$ admits no noiseless subsystems.}

\noindent{{\bf Proof}.  Let us pick a Kraus representation
$\set{K_i}$ of $T$, and let $\frak{K}$ be the corresponding
interaction algebra.  We observe that if any $X \in \cB(\cH)$
belongs to $\frak{K}'$, then $X$ is a fixed point of $T$ on
$\cB(\cH)$.  Indeed,
\begin{equation}
X \in {\frak K}' \Longrightarrow T(X) = \sum_i K_i X K^*_i = X
\sum_i K_i K^*_i.
\end{equation}
Since $T$ is unital, $\sum_i K_i K^*_i = \idty$, and thus $T(X) =
X$.  Now, if $X \in \frak{K}'$, then $X^* \in \frak{K}'$ as well,
which implies that $X_1
\defeq (X + X^*)/2$ and $X_2 \defeq (X-X^*)/2i$ belong to $\frak{K}'$.
Therefore we only need to show that any selfadjoint $X \in
\frak{K}'$ has the form $r\idty$ for some $r \in {\mathbb R}$. For
any selfadjoint $X$, the operator $\abs{X} \defeq (X^2)^{1/2}$
belongs to the algebra generated by $X^2$, so
\begin{equation}
X = X^* \in {\frak K}' \Longrightarrow X_\pm \defeq \frac{\abs{X}
\pm X}{2} \in {\frak K}'.
\end{equation}
Since $X = X_+ - X_-$ and $X_\pm \ge 0$, we reduce our task to
proving that any positive $X$ in $\frak{K}'$ is a multiple of the
identity. Without loss of generality, we may assume that
$\trnorm{X} = 1$. Since $X \ge 0$, we must have $X \in \cS(\cH)$;
since $X$ belongs to the commutant of $\frak{K}$, it is also a
fixed point of $T$.  Thus $X = \idty/\dim \cH$, and the commutant
${\frak K}'$ of the interaction algebra ${\frak K}$ consists of
complex multiples of the identity. \hfill $\blacksquare$}

The proof of Proposition 2 depends in an essential way on the
uniqueness of the fixed point of a strictly contractive channel,
as well as on the condition satisfied by the Kraus operators of a
unital channel. It turns out, however, that the statement of
Proposition 2 can be strengthened to include {\em all} strictly
contractive channels.

\noindent{{\bf Proposition 3}.  Let $T$ be a strictly contractive
channel. Then $T$ admits no noiseless subsystems.}

\noindent{{\bf Proof}.  Let $\frak{K}$ be the interaction algebra
of the channel $T$.  Let us suppose, contrary to the statement of
the Proposition, that $T$ admits at least one noiseless subsystem
(i.e., $\frak{K}$ is reducible).  That is, there exists at least
one $j \in \set{1,\ldots,r}$ such that $m_j,n_j \ge 2$ in
Eqs.~(\ref{eq:intalg1})-(\ref{eq:hsisomorph}).  Let $\cK$ be some
closed subspace of $\cH$.  Restricting the channel $T$ to the set
\begin{equation}
\cS(\cK) \defeq \setcond{\rho \in \cS(\cH)}{{\rm supp}\,\rho
\subseteq \cK}
\end{equation}
(where ${\rm supp}\,\rho$ is the orthogonal complement of $\ker \rho$), we note that, by definition, the contractivity modulus of the
restricted channel cannot exceed the contractivity modulus of $T$.
Let $\cH_j$ be the $j$th direct summand $\tensprod{{\mathbb
C}^{m_j}}{{\mathbb C}^{n_j}}$ in Eq.~(\ref{eq:hsisomorph}). Define
the channel $T_j$ as the restriction of $T$ to $\cS(\cH_j)$. Then
any Kraus operator of $T_j$ has the form
$\tensprod{\idty_{m_j}}{K_\mu}$} where $K_\mu \in M_{n_i}({\mathbb
C})$ and
\begin{equation}
\sum_\mu K^*_\mu K_\mu = \idty_{n_i}.
\end{equation}
Furthermore $\kappa(T_j) \le \kappa(T) < 1$.  Now $T_j$ is the
channel of the form $\tensprod{\id}{S_j}$, where $S_j$ is the
channel on $\cS({\mathbb C}^{n_j})$ with Kraus operators $K_\mu$.
As we have shown in subsection \ref{ssec:defex}, channels of this
form are not strictly contractive (or even weakly contractive).
Thus $\kappa(T_j) = 1$, and the Proposition is proved, {\em
reductio ad absurdum}.\hfill $\blacksquare$

The result of Proposition 3 is quite shocking as it unequivocally
rules out the existence of noiseless subsystems for any strictly
contractive channel.  From the standpoint of foundations of
quantum theory, the importance of Proposition 3 lies in the fact
that it establishes nonexistence of noiseless subsystems for a
wide class of physically realizable quantum computers on the basis
of a minimal set of assumptions.  Furthermore, from the
mathematical point of view, it is rather remarkable that strict
contractivity of a channel already implies irreducibility of its interaction algebra.  We must,
however, hasten to emphasize that, despite its sweeping
generality, Proposition 3 should not be considered as a proof of
impossibility of building a reliable quantum computer.  It merely
rules out the possibility of building quantum computers with {\em
perfect} protection against errors modelled by strictly
contractive channels.

\subsection{Approximate error correction}
\label{ssec:approxec}

At this point we must realize that the results of the previous
subsection are not as unexpected as they may seem.  After all,
nothing is perfect in the real world!  Therefore, our
error-correction schemes must, at best, come as close as possible
to the perfect scenario.  Of course, the precise criteria for
determining how close a given error-correction scheme is to the
"perfect case" will vary depending on the particular situation,
but we can state perhaps the most obvious criterion in terms of
distinguishability of channels.

Let us first phrase everything in abstract terms. Let the error
mechanism affecting the computer be modelled by some channel $T$.
We assume that there exists some positive $\delta < 1$ which, in
some way, characterizes the channel $T$ (it could be given, e.g.,
by the minimum of the operator norms of the Kraus operators of
$T$, and thus quantify the ``smallest'' probability of an error
occurring).  Let $\cH$ be the Hilbert space associated with the
computer.  Then, for each $\epsilon > 0$, we define a {\em
$(\epsilon,\delta)$-approximate error-correcting scheme for $T$}
to consist of the following objects:
\begin{enumerate}
\item[(1)] an integer $n > 1$,
\item[(2)] a Hilbert space $\cH_{\rm ext}$ with $\dim \cH_{\rm ext} \ge \dim
\cH$,
\item[(3)] a channel $\map{E}{\cS(\cH)}{\cS(\cH_{\rm ext})}$,
\item[(4)] a channel $\map{\tilde{T}}{\cS(\cH_{\rm ext})}{\cS(\cH_{\rm ext})}$,
and
\item[(5)] a completely positive map $\map{T_{\rm corr}}{\cS(\cH_{\rm
ext})}{\cS(\cH_{\rm ext})}$,
\end{enumerate}
such that the channel $\tilde{T}$ depends uniquely on $n$,
$\cH_{\rm ext}$, $T$, and $E$; the CP map $T_{\rm corr}$ is correctable
(say, in the Knill-Laflamme sense, or through other means, depending on the particular situation); and we have the estimate
\begin{equation}
\cbnorm{\tilde{T}-T_{\rm corr}} < \delta^n < \epsilon.
\label{eq:approxec}
\end{equation}

Let us give a concrete example in order to illustrate the above
definition. Suppose that the channel $T$ is of the form $\id + S$
with $\cbnorm{S} < \delta$.  Then, for any $n$, we can write
\begin{equation}
T^{\otimes n} = \id + \sum_{A \subset \set{1,\ldots,n} \atop 0<
\abs{A} < n} \bigotimes^n_{k=1} S^{\iota_A(k)} + S^{\otimes n},
\label{eq:excorr1}
\end{equation}
where $\abs{A}$ denotes the cardinality of the set $A$, and
$\map{\iota_A}{\set{1,\ldots,n}}{\set{0,1}}$ is the indicator
function of $A$. We use the convention that, for any map $M$, $M^0
= \id$.  In other words, the summation on the right-hand side of
Eq.~(\ref{eq:excorr1}) consists of tensor product terms with one
or more identity factors.  For the last term, we have
$\cbnorm{S^{\otimes n}} < \delta^n$.

In this case, given some $\epsilon > 0$, we pick such $n$ that
$\delta^n < \epsilon$ and let $\cH_{\rm ext} \defeq \cH^{\otimes
n}$.  If the CP map given by the sum of the first two terms on the
right-hand side of Eq.(\ref{eq:excorr1}) is correctable on some
subspace $\cK$ of $\cH_{\rm ext}$, then the channel $E$ is defined
in a natural way through the composition of the following two
operations:  (a) adjoining additional $n-1$ copies of $\cH$, each
in some suitable state $\rho_0$, and (b) restricting to the
subspace $\cK$. This way, we obviously have $\tilde{T} \defeq
T^{\otimes n}$ and
\begin{equation}
T_{\rm corr} \defeq  \id + \sum_{A \subset \set{1,\ldots,n} \atop
0< \abs{A} < n} \bigotimes^n_{k=1} S^{\iota_A(k)}.
\end{equation}
The estimate (\ref{eq:approxec}) holds because $\tilde{T}-T_{\rm
corr} = S^{\otimes n}$.  We note that this construction results in
a quantum error-correcting code that corrects any $n-1$ errors. We
can use similar reasoning to describe quantum codes that correct
$k < n$ errors.

Constructing $\cH_{\rm ext}$ as a tensor product of a number of
copies of $\cH$, the Hilbert space of the computer, evidently
leads to the usual schemes for fault-tolerant quantum computation
\cite{ftqc}.  Other solutions, such as embedding the
finite-dimensional Hilbert space $\cH$ in a suitable
infinite-dimensional Hilbert space (e.g., encoding a qubit in a
harmonic oscillator \cite{gkp}), can also be formulated in a
manner consistent with our definition above.

Let us now address approximate correctability of strictly
contractive errors.  In subsection \ref{ssec:noisyqmem} we have
demonstrated that, in the absence of error correction, the
sensitivity of quantum memories and computers to such errors grows
exponentially with storage and computation time respectively.  Let
$T$ be a strictly contractive error channel.  It is obvious that
the appropriate approximate error-correction scheme must be such
that the contraction rate of the "encoded" computer, where the
errors are now modelled by the channel $\tilde{T}$, is effectively
slowed down.  In some cases, straightforward tensor-product
realization may prove useful (e.g., when the product channel
$\tensprod{T}{T}$ is not strictly contractive). We must recall
that, for any channel $S$, a necessary condition for
correctability is $\kappa(S) = 1$. Thus, if we can find a suitable
approximate error-correcting scheme where $\tilde{T}$ would be
well approximated by some channel $T_{\rm corr}$ with
$\kappa(T_{\rm corr})=1$, we may effectively slow down the
contraction rate by protecting the encoded computer against errors
modelled by $T_{\rm corr}$.  A more ingenious approach may call
for replacing circuit-based quantum computation with that in
massively parallel arrays of interacting particles; several such
implementations have already been proposed \cite{masspar}.  It is
quite likely that the possible "encodings" of quantum computation
in these massively parallel systems \cite{massparnote} may offer a
more efficient implementation of approximate error correction.

Finally, we should mention that the idea of "approximate" noiseless subsystems has already been explored by Bacon et al. \cite{bacon}.  In their work, it is argued that the symmetry, which is required of a channel in order for noiseless subsystems to exist, is generally broken by perturbing the channel.  They show that, if the perturbations of the channel are "reasonable," then the noiseless subsytem is stable to second order in time.  We must reiterate that the negative results we have stated in the previous subsections refer only to nonexistence of "perfectly" noiseless subsystems; in the real world, we would have no choice but to settle for "almost perfect" anyway.

\section{Conclusion}
\label{sec:conclude}

In this paper, we have offered an argument that errors in
physically realizable quantum computers are naturally modelled by
strictly contractive channels, i.e., channels that uniformly
shrink, in the trace norm, the set of all density operators of the
system under consideration.  In particular, no two density
operators in the image of a strictly contractive channel have
orthogonal supports, which implies that any measurement designed
to distinguish between these density operators will err with
probability bounded away from zero.  This implies, in turn, that
there exists some precision threshold $\epsilon > 0$ such that any
two density operators $\rho,\sigma$ with $\trnorm{\rho-\sigma} <
\epsilon$ cannot be distinguished by a particular experimentally available
measuring apparatus.

We can turn this reasoning around by first postulating the
existence of a precision threshold $\epsilon$ that would quantify
resolving power of the least precise instrument employed in the
experiment.  As we have argued, the physical interpretation of the
precision threshold boils down to limits on our ability to
distinguish between density operators.  A nonzero lower bound on
the probability of error in optimum quantum hypothesis testing can
thus be taken as an indication that the combined influence of
environmental noise and experimental imprecisions (which, in fact,
are quite likely to be caused by indelible quantum-mechanical
effects, such as vacuum fluctuations) can be economically captured
by the concept of a strictly contractive channel.

As we have shown, the set $C_{\rm sc}$ of all strictly contractive
channels on a given system {\bf Q} is dense in the set $C$ of all
channels on {\bf Q}.  Since no finite-precision measurement will
be able to distinguish between an arbitrary channel $T$ and some
strictlly contractive channel $T'$, it is reasonable to ascribe to
strictly contractive channels the property of "experimental
reality," just as we would ascribe this property to elements of
the set $[0,2\pi)\, \cap \, {\mathbb Q}$ (where $\mathbb Q$ is the set of rational numbers) in any experiment
involving finite-precision measurements of angles.

In light of this interpretation, it is important to investigate
the robustness of quantum memories and computers in the presence
of strictly contractive errors.  We have found that, in the
absence of error correction, any state stored in a noisy quantum
register converges exponentially fast to the fixed point of the
error channel $T$, and the rate of convergence is independent of
dimension of the register Hilbert space.  In other words,
sensitivity of quantum registers to strictly contractive noise is
an intensive property, i.e., independent of the register's size.
Similarly, computations performed on a noisy quantum computer with
different initial states quickly yield indistinguishable results,
again at a rate that does not depend on the computer's size
(number of qubits).  Furthermore, the property of strict
contractivity turns out to be strong enough to proscribe the
existence of noiseless subsystems of the computer affected by any
strictly contractive error channel.

However, these results are more of a blessing than a curse for the
future of quantum information processing:  they certainly indicate
that the successful solution of problems faced by researchers in
this field will require models of computers far more ingenious
than networks of one- and two-qubit gates.  As we have mentioned
above, systems of interacting particles (or quantum cellular
automata) may well prove to be a viable medium for experimental
realization of large-scale quantum computers. In this respect, we
would like to point out a possible connection between strictly
contractive channels and ergodic quantum cellular automata
\cite{ricwer}. A cellular automaton is ergodic if it possesses a
unique invariant state which it reaches irrespective of
initial conditions, and this is exactly the property shared by
quantum systems under the influence of strictly contractive
errors.  As an example, let us consider information storage in a
quantum cellular automaton.  It is essential that this automaton
be nonergodic, for otherwise it would not be able to "remember"
anything.  Assuming that each cell (site) of the automaton is
under the influence of some strictly contractive error channel
$T$, an interesting problem would be to devise such a transition
rule that the automaton would not be ergodic.  In this respect, we
should mention that, while $T$ is a strictly contractive channel,
it is not at all obvious whether $\tensprod{T}{T}$ is strictly
contractive as well:  it has a unique fixed point among the
product density operators, namely $\tensprod{\rho_T}{\rho_T}$, but
there may also be another fixed point of $\tensprod{T}{T}$ that is
not a product density operator.

Finally we mention one more point worth exploring.  In our
discussion of physically realizable quantum computers, we have
implicitly assumed that the (im)precision of all experimentally
available procedures can be traced back to the (im)precision of
state preparation, quantified by some threshold value $\epsilon$
(i.e., when we say that state $\rho$ has been prepared, we mean
that any state $\sigma$ with $\trnorm{\rho-\sigma} < \epsilon$ may
have emerged from our preparing apparatus), as well as the
(im)precision of measurements (we would not be able to distinguish
any two states $\rho,\sigma$ with $\trnorm{\rho-\sigma} <
\epsilon$).  This is suggestive of Ludwig's axiomatics of quantum
theory \cite{ludwig}, and it would be theoretically rewarding to
consider physically realizable quantum computers from this
axiomatic perspective as well.

\acknowledgments The author would like to thank:  G.M. D'Ariano, M.
Ozawa, and H.P. Yuen for enlightening discussions and constructive
criticism; D.A. Lidar for interesting comments and pointers to useful references; and the referee for helpful suggestions. This work was supported by the U.S. Army Research
Office through MURI grant DAAD19-00-1-0177.

\end{multicols}


\begin{references}
\bibitem{note1}From this point on, we will refer to quantum computers as simply
"computers" for brevity.
\bibitem{massparnote}The term "massively parallel" here refers to the
symmetrical nature of quantum arrays of interacting particles.
This usage is different from the usual connotation of massive
parallelism, intrinsic in the mathematical structure of quantum
theory.
\end{references}
\end{document}